\documentclass[aps,prc,reprint,twocolumn,superscriptaddress,nofootinbib]{revtex4-2}
\usepackage{graphicx,epsfig}
\usepackage{amsfonts}
\usepackage{amsmath}
\usepackage{amssymb}
\usepackage{braket}
\usepackage{multirow}
\usepackage{commath}
\usepackage{orcidlink}
\usepackage{subfigure}
\usepackage{hyperref}
\hypersetup{colorlinks=true,
            linkcolor=blue,
            anchorcolor=blue,
            citecolor=blue}

\begin{document}
\title{Comparison of variational quantum eigensolvers in light nuclei}

\author{Miquel Carrasco-Codina\orcidlink{0009-0006-1601-322X}}
\email{miquel.carrasco.codina@upc.edu}

\affiliation{
 Departament de Física Quàntica i Astrofísica (FQA), Universitat de Barcelona (UB), c.\ Martí i Franqués, 1, 08028 Barcelona, Spain}
\affiliation{
Institut de Ciències del Cosmos (ICCUB), Universitat de Barcelona (UB), c.\ Martí i Franqués, 1, 08028 Barcelona, Spain}
\affiliation{N3Cat, Universitat Politècnica de Catalunya (UPC), c. Jordi Girona, 1-3, 08034 Barcelona, Spain}

\author{Emanuele Costa
\orcidlink{0000-0002-9830-3119}
}
\email{emanuele.costa@icc.ub.edu}
\affiliation{
 Departament de Física Quàntica i Astrofísica (FQA), Universitat de Barcelona (UB), c.\ Martí i Franqués, 1, 08028 Barcelona, Spain}
\affiliation{
Institut de Ciències del Cosmos (ICCUB), Universitat de Barcelona (UB), c.\ Martí i Franqués, 1, 08028 Barcelona, Spain}

\author{Antonio Márquez Romero
\orcidlink{0000-0002-1435-8017}
}
\email{antonio.marquezromero@fujitsu.com}
\affiliation{
 Departament de Física Quàntica i Astrofísica (FQA), Universitat de Barcelona (UB), c.\ Martí i Franqués, 1, 08028 Barcelona, Spain}
\affiliation{
Institut de Ciències del Cosmos (ICCUB), Universitat de Barcelona (UB), c.\ Martí i Franqués, 1, 08028 Barcelona, Spain}

\author{Javier Men\'{e}ndez 
\orcidlink{0000-0002-1355-4147}
}
\email{menendez@fqa.ub.edu}
\affiliation{
 Departament de Física Quàntica i Astrofísica (FQA), Universitat de Barcelona (UB), c.\ Martí i Franqués, 1, 08028 Barcelona, Spain}
\affiliation{
Institut de Ciències del Cosmos (ICCUB), Universitat de Barcelona (UB), c.\ Martí i Franqués, 1, 08028 Barcelona, Spain}

\author{Arnau Rios 
\orcidlink{0000-0002-8759-3202}}
\email{arnau.rios@fqa.ub.edu}
\affiliation{
 Departament de Física Quàntica i Astrofísica (FQA), Universitat de Barcelona (UB), c.\ Martí i Franqués, 1, 08028 Barcelona, Spain}
\affiliation{
Institut de Ciències del Cosmos (ICCUB), Universitat de Barcelona (UB), c.\ Martí i Franqués, 1, 08028 Barcelona, Spain}

\begin{abstract}
Quantum computing is one of the most promising technologies of the near future, and the simulation of quantum many-body systems is a natural application. In this work, we present classical simulations of the ground states of light atomic nuclei within the $p$ shell, from $^{6}$He to $^{10}$B, calculated within the nuclear shell model. We compare the performance of two leading variational quantum eigensolver algorithms: the Unitary Coupled Cluster (UCC) and the Adaptive Derivative-Assembled Pseudo-Trotter (ADAPT) methods, introducing a new metric to quantify the use of quantum resources in each simulation. 
We find that Slater determinants are the most useful reference states for both approaches. 
Our analysis suggests that ADAPT is more efficient for nuclei close to magic numbers, while UCC tends to require fewer resources toward the mid shell. This work lays the groundwork for robust benchmarking of quantum algorithms in nuclear structure studies. 
\end{abstract}

\maketitle

\section{Introduction}
The simulation of quantum many-body systems in classical devices is challenged by the exponential growth of the dimension of the Hilbert space and the entanglement correlations between particles~\cite{Feynman1982-dg}. For this reason, the study of nuclear structure is limited by a lack of precise simulations, especially in heavy nuclei. The development of new techniques to address these systems can enhance the understanding of strong~\cite{Strongforce,Shapes1,Shapes2, Shapes3,Strongforce2,Strongforce3,Shapes4} and weak~\cite{Weakforce,Weakforce2} interactions in nuclei, as well as phenomena such as entanglement correlations among nucleons~\cite{Robin:2020aeh,Johnson:2022mzk,Tichai:2022bxr,Entanglement_ICCUB,Brokemeier:2024lhq}. Moreover, accurate simulations of nuclear structure are essential to tackle fundamental questions in astrophysics~\cite{Originelements,darkmatter,Neutronstars} and particle physics~\cite{Neutrinos2, Standmodel}.

Quantum computing offers an alternative to classical simulations for solving many-body problems, exploiting the quantum-mechanical behavior of its fundamental elements, qubits~\cite{Nielsen_Chuang_2010, DiVincenzo_2000}. The concept of a computer that simulates complex many-body systems by encoding information in quantum particles has been established for decades~\cite{Feynman1982-dg}, but interest in this approach has surged recently due to rapid advancements in quantum-computing devices~\cite{Preskill_2018}.  
However, the inherently sensitive nature of qubits still causes several limitations~\cite{Preskill_2018, Noise, Cryo}. 
These challenges have driven the development of hybrid algorithms, which optimize computations by distributing tasks between quantum and classical devices, thus mitigating the high error rates associated with quantum circuits in the current noisy intermediate-scale quantum (NISQ) era. 

Various quantum approaches have been proposed to study nuclear structure~\cite{Ayral2023}. Some algorithms rely on projection techniques~\cite{Stetcu2023,Li2024,Rule2024,Yoshida:2024ubi} while others propose annealing protocols~\cite{Du:2021ctr,Costa:2024ede}. Many works focus on hybrid algorithms for many-body simulations called Variational Quantum Eigensolvers (VQEs)~\cite{1st_VQE, All_vqe, Review_VQE}, initially using the Unitary Coupled Cluster (UCC) approach~\cite{Dumitrescu2018,Oriel_kiss, Stetcu2022,Sarma2023}.
More recently, also the Adaptive Derivative-Assembled Pseudo-Trotter (ADAPT)~\cite{Romero:2022blx,ADAPT_ICCUB,Zhang:2024uxp,Perez-Obiol:2024vjo,Lacroix2025,Singh2025} and other problem-inspired, hardware-efficient algorithms have been employed~\cite{Bhoy2024,Singh2025}. 
Promising developmemts related to magic and non-stabilizerness have been recently explored in nuclear systems too~\cite{Brokemeier:2024lhq,Robin2025}.
VQEs are especially appealing since they have been applied to a variety of many-body systems, including quantum chemistry~\cite{ADAPT, QuantChemis, Quantum_chem}, Ising~\cite{Ising} and Lipkin-Meshkov-Glick models~\cite{Lipkin1, Lipkin2, Lipkin3, Lipkin4, Baid2024}, and superfluids~\cite{Superfluids1, Superfluids2}. In the context of nuclear structure, while various strategies within the same VQE framework have been compared in detail for UCC~\cite{Oriel_kiss}, an equal-footing evaluation of the performance of various VQEs applied to the same nuclei is still missing.

In this work, we employ classical simulations to analyze the performance in solving nuclear ground states of two state-of-the-art VQEs: UCC~\cite{UCC} and ADAPT~\cite{ADAPT}. We apply these algorithms to the same nuclei within a common many-body framework: the nuclear shell model (NSM). Our aim is to evaluate the quantum resources for each method, thereby identifying the best-suited algorithm for each NSM simulations. To achieve this, we introduce a new metric based on the number of quantum operations required to simulate a given system, comparing the performance of both VQEs across isotopes in the $p$ shell. Our comparisons also employ different reference states and randomized initial parameters, in an effort to obtain the most general possible method-independent conclusions. This approach provides us with unique insights into the suitability of quantum algorithms for studying nuclear structure.

\begin{figure*}[t!]
    \centering
    \subfigure{
        \includegraphics[width=0.45\textwidth]{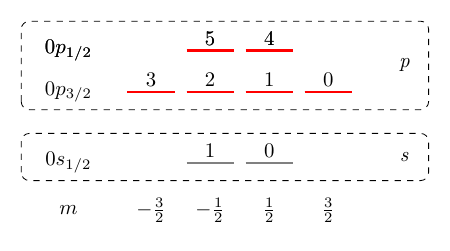}
        \label{fig:shells}}
    \hspace{0\textwidth}
    \subfigure{
        \includegraphics[width=0.35\textwidth]{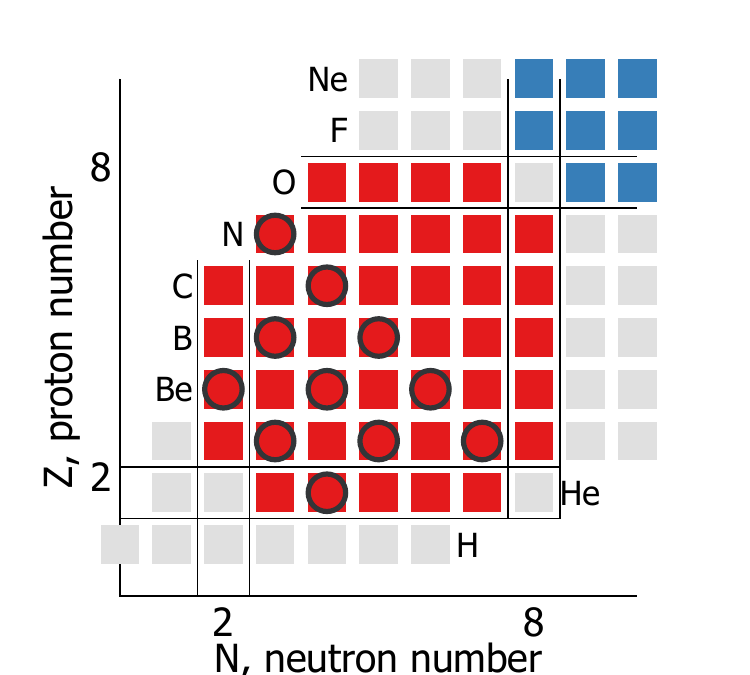}
        \label{fig:pshell_squares}}
    \caption{Left panel: Single-particle states in the lowest nuclear shells. Each state is labeled by its quantum numbers $nl_{j}$ ($y$ axis) and $m$ ($x$ axis). The index corresponds to a qubit in the VQE circuit, according to a Jordan-Wigner mapping~\cite{JW}. Right panel: Nuclear chart for the lightest nuclei, based on the number of neutrons, $N$, ($x$ axis) and the number of protons, $Z$, ($y$ axis). The $p$ shell is displayed in red and nuclei simulated in this work are highlighted with grey circles.}
    \label{fig:two_images}
\end{figure*}

This paper is structured as follows: Sec.~\ref{sec:NSM} offers a brief introduction to the NSM and presents the nuclei studied in our work. We then discuss the details of the UCC and ADAPT methods in Sec.~\ref{sec:VQE}. The results of our comparison are provided in Sec.~\ref{sec:Results}. Finally, we draw our conclusions and outline future research in Sec.~\ref{sec:Conclusions}.

\section{Nuclear Shell Model}
\label{sec:NSM}
The NSM describes nuclear structure based on a valence space comprised of orbitals with similar energies that form a \emph{shell}. These shells contain quantum single-particle states for protons and neutrons, the constituents of the nucleus. When fully occupied, shells lead to the so-called \emph{magic numbers}~\cite{Mayer_II,Jensen}, which are associated to large energy gaps between orbitals. Within the NSM~\cite{Shell_model_hamiltonian}, the complexity of full nuclear systems is simplified by assuming that nucleons in the valence space can describe all the nuclear structure of
each isotope in a given shell. Active nucleons in the valence space are decoupled from the frozen degrees of freedom in the core~\cite{Effective_Interaction,Interactions_Nuclear}. 

The single-particle states of the valence space are defined by the principal quantum number, $n$; the orbital angular momentum, $l$; the total angular momentum, $j$, along with its projection, $m$; and the isospin projections for protons, $t_{z}=1/2$, and neutrons, $t_{z}=-1/2$. 

The left panel of Fig.~\ref{fig:two_images} shows the single-particle states within the $s$ and $p$ shells, according to $n$, $l$, $j$, and $m$. In this work, we consider nuclei with valence-space nucleons in the $p$ shell, with the $s$ shell as core.

In second quantization, the NSM Hamiltonian is written as
\begin{align}
    \hat{H}_\text{eff}=
    \sum_{i} \varepsilon_{i } \, \hat{a}^{\dagger}_{i} \hat{a}_{i} 
    + \frac{1}{4} \sum_{i j k l} \overline{v}_{i j k l} \, \hat{a}^{\dagger}_{i} \hat{a}^{\dagger}_{ j } \hat{a}_{ l }\hat{a}_{k},
\end{align}
where $\hat{a}^{\dagger}_i$ and $\hat{a}_i$ are the fermionic creation and annihilation operators for a state $i$, obeying the anticommutation relations
\begin{align}
    \{\hat{a}^{\dagger}_{i },\hat{a}^{\dagger}_{ j }\}=\{\hat{a}_{ i },\hat{a}_{j }\}=0, \quad 
    \{\hat{a}^{\dagger}_{i },\hat{a}_{j}\}=\delta_{i j} .
\end{align}
$\varepsilon_{i}$ represents the single-particle energy of the $i$-th state, that collectively includes quantum numbers, $i=(n, l, j, m, t_z)$, and $\overline{v}_{i j k l}$ are the antisymmetrised two-body interaction matrix elements. We use the standard Cohen-Kurath interaction~\cite{Cohen_Kurath} as $\hat{H}_\text{eff}$ in the $p$ shell.

In this setting, the many-body basis consists of Slater determinants~\cite{Slater}. Every nuclear shell with a set of orbitals with $j$ quantum number has single-particle dimension
\begin{align}
    \mathcal{D}_\text{sh} = \sum_j (2j + 1) \, ,
\end{align}
for both protons and neutrons. 
The dimension of the corresponding  many-body Hilbert space is, 
\begin{align}
    \text{dim} ( \mathcal{H} )= 
    \binom{ \mathcal{D}_\text{sh} }{ N_\text{val} }
    \binom{ \mathcal{D}_\text{sh} }{ Z_\text{val} } \, ,
\end{align}
where $N_\text{val}$ ($Z_\text{val}$) is the corresponding valence neutron (proton) number. This quantity grows combinatorially with the number of particles, creating a bottleneck for classical simulations. 

Our work aims to benchmark two computationally-intensive VQEs. Beyond quantifying the required resources, we seek to assess the potential of these methods for studying nuclear structure. To this end, we explore several nuclei within a small Hilbert space, the $p$ shell, where we can solve the many-body problem through direct diagonalization.  Table~\ref{tab: all_nuclei} lists the eleven nuclei that we have simulated, highlighted with gray circles in the nuclear chart in the right panel of Fig.~\ref{fig:two_images}. All systems contain an even total number of nucleons, including both even-even and odd-odd nuclei. Column $6$ in Table~\ref{tab: all_nuclei} gives the many-body dimension for the selected $p$-shell isotopes, ranging from $\text{dim} (\mathcal{H})=5$ for systems with two nucleons on top of the core ($^6$Be and $^6$He) to $\text{dim} (\mathcal{H})=84$ for the mid-shell nucleus $^{10}$B. We denote each state of the many-body basis (or Slater determinant) as $\ket{ v_\alpha }$, with $\alpha=0, \cdots, \text{dim} (\mathcal{H})-1$. 

Since $\hat{H}_\text{eff}$ is rotationally invariant, the entire system has well-defined total angular momentum, $J$, and total magnetic number, $M$. We work with the uncoupled basis and restrict the many-body basis according to $M$. Since the number of nucleons in all nuclei in Table~\ref{tab: all_nuclei} is even, all ground states have $M=0$. We exploit this property and restrict our simulations by starting from reference states with $M=0$, excluding excitation operators beyond this subspace~\cite{ADAPT_ICCUB,Bhoy2024}. Alternative projection techniques, such as the one proposed by Ref.~\cite{Rule2024}, could also be employed.

\begin{table}[]
\begin{tabular}{lcccccc}
\hline \hline
\multicolumn{1}{l|}{Nucleus} & \multicolumn{1}{c|}{$Z$} & \multicolumn{1}{c|}{$N$} & \multicolumn{1}{c|}{$Z_{\rm{val}}$} &\multicolumn{1}{c|}{$N_{\rm{val}}$} & \multicolumn{1}{c|}{$\dim(\mathcal{H})$}              & $\Delta$        \\ \hline
\multicolumn{1}{c|}{$^{6}$He}     & \multicolumn{1}{c|}{2} & \multicolumn{1}{c|}{4} & \multicolumn{1}{c|}{0} & \multicolumn{1}{c|}{2} & \multicolumn{1}{c|}{\multirow{2}{*}{5}}  & \multirow{2}{*}{8}   \\
\multicolumn{1}{c|}{$^{6}$Be}     & \multicolumn{1}{c|}{4} & \multicolumn{1}{c|}{2} & \multicolumn{1}{c|}{2} & \multicolumn{1}{c|}{0} & \multicolumn{1}{c|}{}                    &                      \\ \hline
\multicolumn{1}{c|}{$^{6}$Li}     & \multicolumn{1}{c|}{3} & \multicolumn{1}{c|}{3} & \multicolumn{1}{c|}{1} & \multicolumn{1}{c|}{1} & \multicolumn{1}{c|}{10}                  & 45                   \\ \hline
\multicolumn{1}{c|}{$^{8}$Li}     & \multicolumn{1}{c|}{3} & \multicolumn{1}{c|}{5} & \multicolumn{1}{c|}{1} & \multicolumn{1}{c|}{3} & \multicolumn{1}{c|}{\multirow{2}{*}{28}} & \multirow{2}{*}{129} \\
\multicolumn{1}{c|}{$^{8}$B}      & \multicolumn{1}{c|}{5} & \multicolumn{1}{c|}{3} & \multicolumn{1}{c|}{3} & \multicolumn{1}{c|}{1} & \multicolumn{1}{c|}{}                    &                      \\ \hline
\multicolumn{1}{c|}{$^{8}$Be}     & \multicolumn{1}{c|}{4} & \multicolumn{1}{c|}{4} & \multicolumn{1}{c|}{2} & \multicolumn{1}{c|}{2} & \multicolumn{1}{c|}{51}                  & 145                  \\ \hline
\multicolumn{1}{c|}{$^{10}$Li}    & \multicolumn{1}{c|}{3} & \multicolumn{1}{c|}{7} & \multicolumn{1}{c|}{1} & \multicolumn{1}{c|}{5} & \multicolumn{1}{c|}{\multirow{2}{*}{10}} & \multirow{2}{*}{49}  \\
\multicolumn{1}{c|}{$^{10}$N}     & \multicolumn{1}{c|}{7} & \multicolumn{1}{c|}{3} & \multicolumn{1}{c|}{5} & \multicolumn{1}{c|}{1} & \multicolumn{1}{c|}{}                    &                      \\ \hline
\multicolumn{1}{c|}{$^{10}$Be}    & \multicolumn{1}{c|}{4} & \multicolumn{1}{c|}{6} & \multicolumn{1}{c|}{2} & \multicolumn{1}{c|}{4} & \multicolumn{1}{c|}{\multirow{2}{*}{51}} & \multirow{2}{*}{145} \\
\multicolumn{1}{c|}{$^{10}$C}     & \multicolumn{1}{c|}{6} & \multicolumn{1}{c|}{4} & \multicolumn{1}{c|}{4} & \multicolumn{1}{c|}{2} & \multicolumn{1}{c|}{}                    &                      \\ \hline
\multicolumn{1}{c|}{$^{10}$B}     & \multicolumn{1}{c|}{5} & \multicolumn{1}{c|}{5} & \multicolumn{1}{c|}{3} & \multicolumn{1}{c|}{3} & \multicolumn{1}{c|}{84}                  & 145                  \\ \hline \hline
\end{tabular}
\caption{Nuclei simulated in this work using the UCC and ADAPT methods. The table gives the total number of protons, $Z$; neutrons, $N$; the number of valence protons, $Z_{\rm{val}}$; and valence neutrons, $N_{\rm{val}}$; as well as the number of Slater determinants, $\text{dim} (\mathcal{H})$; and the number of pool operators, $\Delta$, in the $p$ shell.}
\label{tab: all_nuclei}
\end{table}

\section{Variational Quantum Eigensolvers}
\label{sec:VQE}
VQEs take advantage of the Rayleigh-Ritz variational principle~\cite{Ritz1909} by preparing a parametrized quantum state, the \emph{ansatz} $\ket{\psi (\boldsymbol{\theta})}$, in a quantum device. The energy is minimized with respect to the \emph{ansatz} parameters, $\boldsymbol{\theta}=\{\theta_0,\theta_1,...,\theta_R\}$, where $R$ represents the dimension of the energy landscape, $E(\boldsymbol{\theta})=\bra{\psi(\boldsymbol{\theta})}\hat{H}_\text{eff} \ket{\psi(\boldsymbol{\theta})}$. The ground state search is a classical optimization problem, with energies obtained by performing adequate measurements on the quantum state~\cite{ADAPT_ICCUB}. Thus, VQEs exploit the quantum-mechanical properties of qubits to reproduce quantum states while performing most computations on the classical optimizer. 

The structure of the energy landscape in many-body simulations may be particularly challenging, either because of the presence of multiple local minima~\cite{Larocca2022diagnosingbarren} or restrictions that exclude sectors of the Hilbert space~\cite{Review_VQE}.
Therefore, to converge to the system's ground state, the \emph{ansatz} has to be carefully prepared, allowing the parametrized circuit to be expressive enough to explore the entire configuration space~\cite{Lie_algebra, All_vqe}. In the following subsections, we discuss the state preparation employed by UCC and ADAPT before addressing a full comparison of the two methods. 

\subsection{Unitary Coupled Cluster}\label{sec:UCC}

The UCC approach is inspired by the success of classical Coupled Cluster methods to solve the quantum many-body problem~\cite{Coupled_Cluster}, by arranging an \emph{ansatz} made of unitary operators that can be prepared in a quantum computer. The UCC \emph{ansatz} we consider in this work consists of a reference state and a product of unitary operators, 
\begin{align}
\ket{\psi_{\rm{UCC}}(\boldsymbol{\theta})}=\prod_{ijkl}e^{\theta_{ij}^{kl}\hat{T}_{ij}^{kl}}\ket{\psi_{0}},
    \label{Eq:ansatz_UCC}
\end{align}
where $\theta_{ij}^{kl}$ are variational parameters. 
The reference state has an energy 
$E_0= \braket{ \psi_{0} | \hat H_\text{eff} | \psi_{0} } / \braket{ \psi_0 | \psi_0 }$.
In principle, the operators $\hat T$ can be arbitrarily complex in terms of many-body excitation order. We choose two-body fermionic operators $\hat{T}_{ij}^{kl}$ of the form 
\begin{equation}
    \hat{T}_{ij}^{kl} = \hat{a}_{i}^{\dagger}\hat{a}_{j}^{\dagger}\hat{a}_{l}\hat{a}_{k} - \hat{a}_{k}^{\dagger}\hat{a}_{l}^{\dagger}\hat{a}_{j}\hat{a}_{i}\,,
\label{eq:antiherm}
\end{equation}
which are then translated onto qubit operators by means of a specific encoding, like the Jordan-Wigner mapping~\cite{JW}. Given the anti-Hermitian form of the excitation operators in Eq.~\eqref{eq:antiherm},
the exponential operators are unitary and can thus be implemented in quantum circuits using unitary quantum gates~\cite{DiVincenzo_2000, Nielsen_Chuang_2010}.
With the choice of these operators, we provide an \emph{ansatz} that is expressive enough to cover the Hilbert space~\cite{UCC,Romero:2022blx}. This \emph{ansatz} is usually referred to in the literature as UCCD~\cite{sokolov, Romero_ucc}, as it only involves the sum of all \emph{doubles} (D) excitations on top of the reference state in the operator exponent. Moreover, the product form in Eq.~(\ref{Eq:ansatz_UCC}) is obtained after applying the Trotter-Suzuki expansion~\cite{Suzuki1976-mz, Trotter} 
\begin{equation}
    e^{\hat{A}+\hat{B}} = \lim_{n\to\infty} \left(e^{\hat{A}/n}e^{\hat{B}/n}\right)^n + O(1/n),
\end{equation}
with only one step, $n=1$, which has been shown empirically to be enough to yield accurate results~\cite{barkoutsos_ucc, grimsley_ucc}. 

For a given nuclear system, and for a specific type of operators, like the two-body operators of Eq.~\eqref{eq:antiherm}, 
the complete set of available operators is fixed from the outset of a simulation and constitutes the so-called \emph{operator pool}. 
 We refer to the exponential operators as \emph{layers}, each having one excitation, $\hat{T}_{ij}^{kl}$, and one parameter, $\theta_{ij}^{kl}$. The number of  operators included in the \emph{ansatz}, $\Delta$, determines the number of layers and the dimension of the energy landscape, $R=\Delta$.

The size of the UCC operator pool, $\Delta$, is a relevant hyperparameter of the algorithm. While we may want to build a very expressive \emph{ansatz} by increasing the operator pool size, this strategy may not be effective since a richer \emph{ansatz} requires deeper quantum circuits, which are more prone to errors when simulated in a device. Besides, the classical minimization solver benefits from a lower energy-landscape dimension in terms of the parameters describing the \emph{ansatz} and requires fewer evaluations from the quantum device. In that sense, our approach is to build a UCC \emph{ansatz} that can reach as many configurations as possible while eliminating redundant operators.

In order to achieve this, we start with the entire pool of two-body excitation operators, $\hat{T}_{ij}^{kl}$, available within the Hilbert space of the system. We then exclude the operators that annihilate and create the same single-particle states, $\hat{T}_{ij}^{ij}=0$, according to Eq.~(\ref{eq:antiherm}). We also remove the permutations for each set of indices $\{i,j,k,l\}$ that lead to an equivalent operator, e.g. $\hat{T}_{ij}^{kl}=-\hat{T}_{ji}^{kl}=-\hat{T}_{ij}^{lk}$. Finally, we also exclude excitations that do not respect angular-momentum selection rules. Overall, this approach reduces $\Delta$ by a factor of two for all studied nuclei. Column $7$ of Table~\ref{tab: all_nuclei} lists the final operator pool size for all nuclei studied in this work.

While the operator pool is chosen \emph{a priori},  the UCC \emph{ansatz} does not prescribe the ordering of the operators in the product of Eq.~\eqref{Eq:ansatz_UCC}. However, the performance of the method depends on such ordering~\cite{UCC,Oriel_kiss} because, in general, excitation operators do not commute, $[\hat{T}_{ij}^{kl}, \hat{T}_{ab}^{cd}] \neq 0$. Each ordering of operators may produce a distinct quantum  state, leading to variations in the expectation value of the energy and the corresponding variational landscape. To account for this, we consider several orderings for each UCC minimization, specified in Sec.~\ref{sec:Results}.With this generic approach, we aim to have a general overview of the performance of the UCC method, as opposed to focusing on specific minimization instances. 

There are additional factors that influence the minimization process in UCC, including the reference state $\ket{\psi_{0}}$ and the initial parameters $\boldsymbol{\theta}_{0}$. We discuss these aspects, which also impact ADAPT, in Sec.~\ref{sec:Comparisons}.

\subsection{ADAPT}
The ADAPT method was originally developed to reduce computations in quantum devices compared to UCC~\cite{ADAPT}. Unlike UCC, the ADAPT \emph{ansatz} is not fixed upfront but grows iteratively. Starting from a reference state $\ket{\psi_0}$, the \emph{ansatz} is built by application of selected excitation operators, ending up with a similar product structure as in Eq.~\eqref{Eq:ansatz_UCC}. For a given layer, the operator is selected from the same operator pool used for UCC, based on the highest gradient at the origin,
\begin{align}
    \frac{\partial E}{\partial \theta_{ij}^{kl}}\Bigg|_{\theta_{ij}^{kl}=0}=
    \braket{ \psi_0 | [\hat{H}_\text{eff},\hat{T}_{ij}^{kl}]|
    \psi_0 },
    \label{eq:gradient}
\end{align}
aiming for a ``large'' step in the minimization process. 
The optimal parameters are then found classically by minimizing the energy at every iteration.

In nuclei, the quantum resources of the ADAPT algorithm have been discussed in Ref.~\cite{ADAPT_ICCUB}. We note that the gradients can be computed by performing specific adequate measurements on the quantum circuit that parametrizes the \emph{ansatz}~\cite{ADAPT_ICCUB}. 

\subsection{UCC vs ADAPT}
\label{sec:Comparisons}

While the ADAPT \emph{ansatz} is similar to that of UCC, the two methods significantly differ in their minimization strategies. UCC has a $\Delta$-dimensional energy surface from the outset, thus requiring minimizations in large-dimensional spaces. In contrast, ADAPT grows the \emph{ansatz} iteratively on a layer-by-layer basis, potentially requiring fewer resources in nuclei close to magic numbers, where few layers are sufficient to converge the ground state~\cite{ADAPT_ICCUB}. However, in general it is not easy to find a straightforward comparison of the classical and quantum computational complexity of the two methods.

\subsubsection{Total operations}

Standard metrics to evaluate the efficiency of a VQE based on parameter optimization include the number of iterations of the classical minimizer and the number of energy function evaluations~\cite{Oriel_kiss}. However, these quantities do not necessarily reflect how many operations the quantum device performs. Due to the iterative growth of the ADAPT \emph{ansatz}, its initial iterations are significantly less costly for the quantum computer than later ones. Furthermore, the UCC method has a fixed $\Delta$. This requires more gates on the quantum circuit, leading to increased errors compared to the preparation of the first iteration of the ADAPT with a single operator.

We introduce a new metric, \emph{total operations}, that considers the size of the \emph{ansatz} and approximates the number of quantum operations throughout the entire algorithm. It is computed by counting the number of unitary operators that act on the reference state during the VQE. For UCC, the number of total operations is the product of the number of times the optimizer calls the energy function, $\rm{fcalls}$, and the number of operators (layers) in the \emph{ansatz},
\begin{align}
    N_\text{op}^\text{UCC} = \rm{fcalls} \times \Delta \,.
    \label{eq:Nops_UCC}
\end{align}
In contrast, for ADAPT, the total operations metric sums, for each iteration, the number of function calls times the number of operators included in the \emph{ansatz}:
\begin{align}
        N_\text{op}^\text{ADAPT} = \sum_{i=1}^{N_\text{iter}}\rm{fcalls}_{i} \times i. 
        \label{eq:Nops_ADAPT}
\end{align}
It is important to note that, in our results, $N_\text{op}^\text{ADAPT}$ does not include the operations corresponding to the estimation of the gradients of the operators, which is  orders of magnitude smaller than the operations needed for the minimization of the energy~\cite{ADAPT_ICCUB}.

This metric enables a consistent comparison between the two algorithms and provides insight into the number of quantum gates used during VQE minimization. Note that our definition of total operations assumes that each layer can be implemented with the same number of quantum gates in both approaches, which may not always be the case, as gate implementation varies depending on connectivity, hardware, or error mitigation protocols. Nonetheless, we consider $N_\text{op}$ as a faithful representation of the computational load of the quantum device, more accurate than the number of iterations of the optimization method or the number of function calls.

\subsubsection{Reference states and initial parameters}

Some additional aspects need to be addressed before we can directly compare UCC and ADAPT. First, the reference state in Eq.~\eqref{Eq:ansatz_UCC} may play an important role in the minimization. Past implementations for nuclear ground states propose Hartree-Fock solutions as reference states~\cite{Stetcu2022} or many-body basis states of minimal energy~\cite{Romero:2022blx,ADAPT_ICCUB}. One might expect that a reference state that is ``closer'' by some measure to the ground state may converge with fewer quantum resources. We assess this aspect by employing different many-body basis states---Slater determinants---, $\ket{ v_\alpha }$, as initial states for both ADAPT and UCC.  Under the Jordan-Wigner mapping, these states require minimal circuit resources~\cite{ADAPT_ICCUB,Costa:2024ede}. Additionally, we also consider initial states that are linear combinations of Slater determinants,
\begin{equation}
    \ket{  v_\text{rand} }= \sum_\alpha c_\alpha \ket{ v_\alpha },
\end{equation}
with random coefficients $c_\alpha$~\cite{Carrasco}. This allows us to explore the importance of generic reference states across the two VQEs of interest.

Second, the choice of starting parameters in ADAPT is prescribed by the method itself. In minimizing the $l$-th layer, one picks the optimal set of parameters $\boldsymbol{\theta}^*$ of the previous iteration and starts the $l$-th layer at $\theta_l=0$. The minimization may alter the value of all these parameters, although one may keep subsets of them fixed for a few iterations to simplify the minimization procedure~\cite{Perez-Obiol:2024vjo,Lacroix2025}. 

In contrast, the starting parameters of the UCC \emph{ansatz} in Eq.~\eqref{Eq:ansatz_UCC} are not fixed, and different choices may affect the minimization performance. 
We account for this by taking randomized values of the initial parameters, from a uniform distribution between $-\pi$ and $\pi$. This randomization of initial parameters should not be confused with the randomized operator orderings which we also explore in the following section.

\section{Results}
\label{sec:Results}

For all nuclei in Table  \ref{tab: all_nuclei}, we perform  classical UCC and ADAPT simulations using \texttt{Python}. In particular, we use \texttt{NumPy} base array operations to prepare the \emph{ans\"atze}~\cite{NumPy} and \texttt{SciPy}'s optimization functions~\cite{SciPy}, with the L-BFGS-B method~\cite{LBFGSB}. We employ a common tolerance $f_{tol}=10^{-4}$ to minimize the energy functions of both UCC and ADAPT. Our simulations do not include noise. We provide an open code for the implementation of the results in the GitHub repository~\cite{Github}.

To benchmark the convergence of UCC and ADAPT, we compare the VQE results with the ground-state energies obtained from a classical diagonalization of the Hamiltonian for each system, computed using the Lanczos algorithm~\cite{Shell_model_hamiltonian}. We consider a minimization to be successful when the relative error between the ground-state energy and the VQE result satisfies
\begin{equation}
    \varepsilon=\abs{\frac{E_{\rm{VQE}}-E_{\rm{NSM}}}{E_{\rm{NSM}}}}\leq 10^{-4}.
\end{equation}

Using the aforementioned tools, we compare the UCC and ADAPT methods to determine which one is more efficient for simulating nuclear structure on a quantum device. ADAPT minimizations correspond to a single run for each initial reference state. 
In contrast, our UCC results provide central values and standard deviations from multiple minimizations of the UCC \emph{ansatz}. In each UCC run, we pick a random ordering of the operators, which remains fixed across the minimization, as well as a random initial set of parameters.
The number of minimizations we average depends on the dimensions of the Hilbert space of the system, given in Table~\ref{tab: all_nuclei}. We employ $100$ samples for nuclei with $\rm{dim}(\mathcal{H})=5$; $50$ samples if $\rm{dim}(\mathcal{H})=10\text{ or } \rm{dim}(\mathcal{H})=28$; and $30$ samples in isotopes with $\rm{dim}(\mathcal{H})=51\text{ or }\rm{dim}(\mathcal{H})=84$.

\begin{figure*}
\centering
\begin{tabular}{cc}
    \multicolumn{2}{c}{\includegraphics[width=\textwidth]{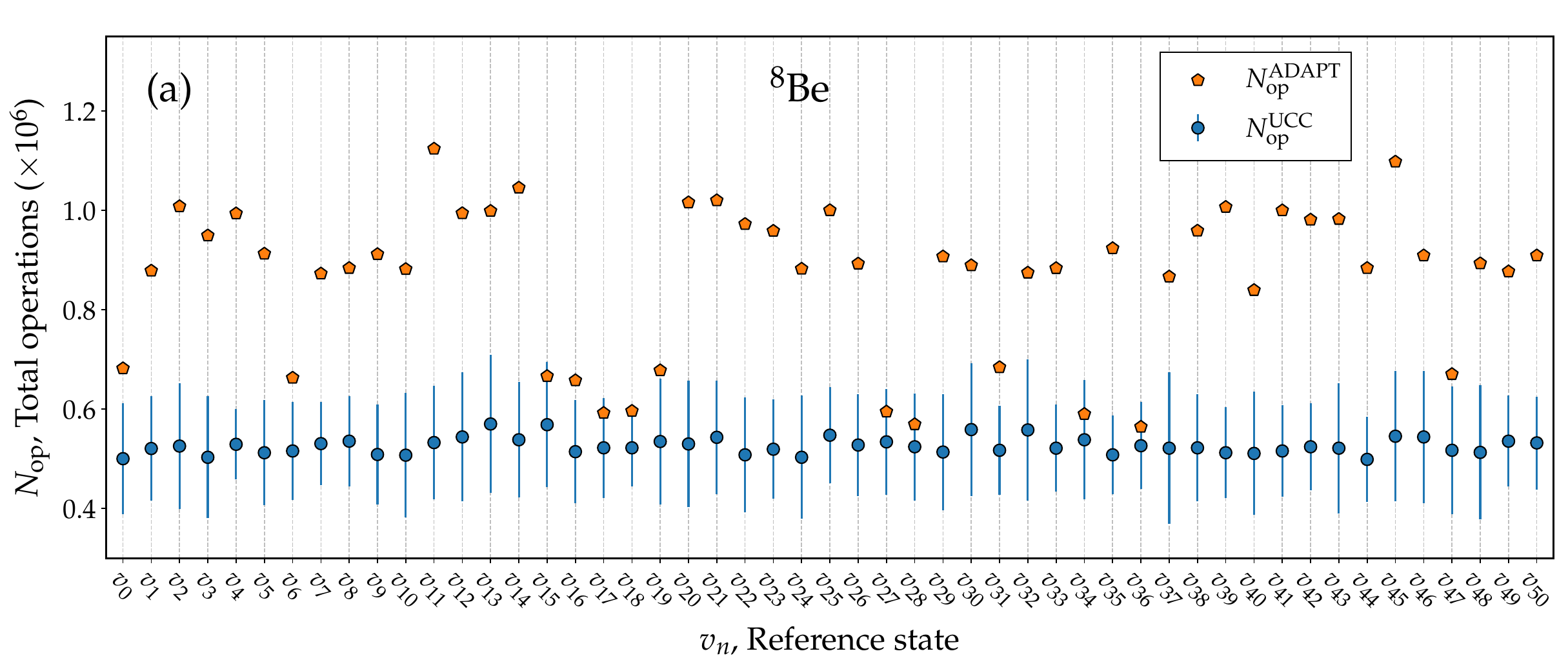} }\\
  \includegraphics[width=0.6666\textwidth]{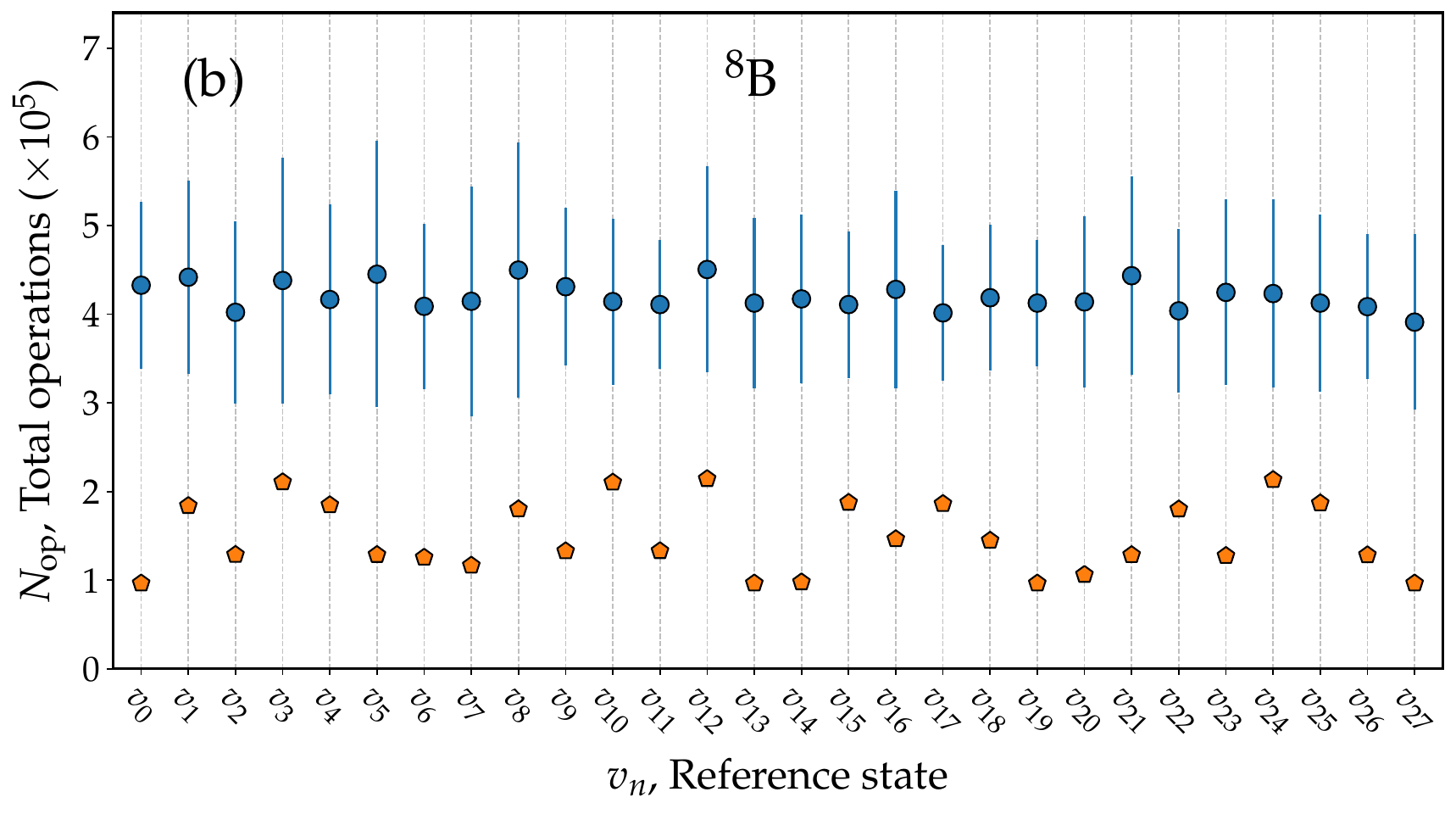} &   \includegraphics[width=0.3333\textwidth]{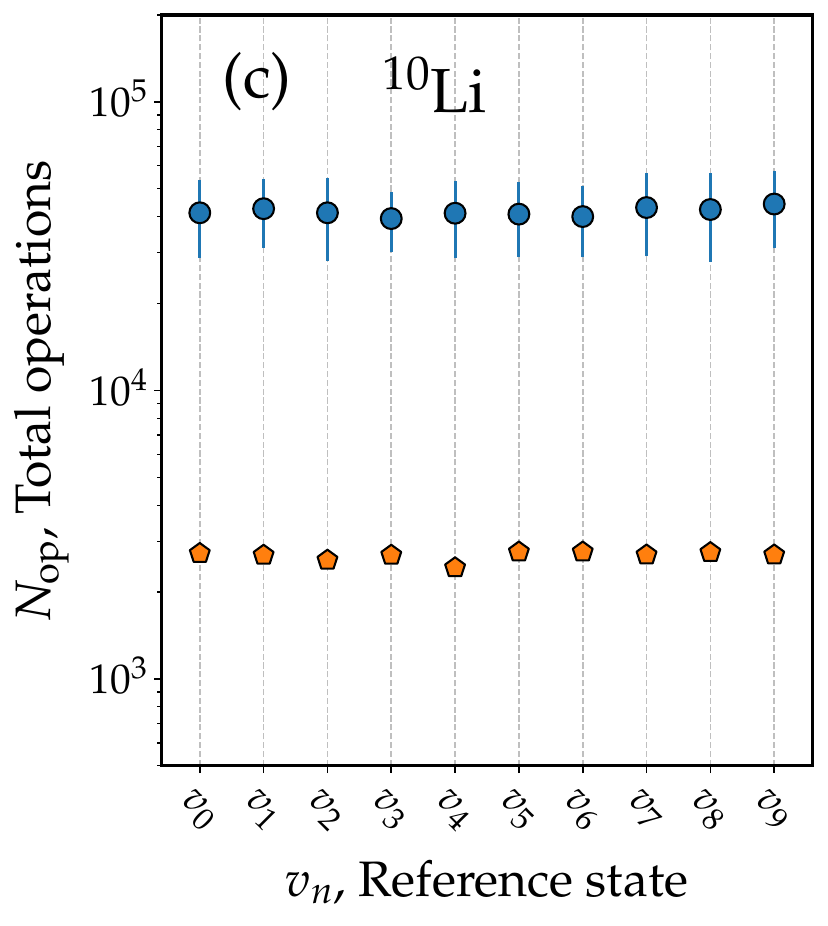} \\
\end{tabular}
\caption{Total operations, $N_{\text{op}}$, needed to converge the ground state of (a) $^{8}$Be, (b) $^{8}$B and (c) $^{10}$Li for all Slater determinants considered in the many-body basis as reference state. The results are quantified using Eq.~\eqref{eq:Nops_ADAPT} for ADAPT (orange pentagons) and Eq.~\eqref{eq:Nops_UCC} for the average (blue circles) and standard deviations (blue bars) of the UCC values (see text for details).}
\label{fig:v_perf_UCC_vs_ADAPT}
\end{figure*}

\subsection{$^8$Be, $^{8}$B and $^{10}$Li: role of the reference state}

We first assess the importance of the reference state for the performance of the two VQEs. This dependence is important to determine method-agnostic (as opposed to state-dependent) comparisons, because the choice of reference state affects the minimization and can therefore impact the efficiency of the whole VQE. 

Figure \ref{fig:v_perf_UCC_vs_ADAPT} compares the total operations for UCC (blue circles) and ADAPT (orange pentagons) in $^{8}$Be (panel a), $^{8}$B (panel b) and $^{10}$Li (panel c). These three nuclei cover a wide range of configuration-space dimensions  ($\dim{(\mathcal{H})}=51$, $\dim{(\mathcal{H})}=28$ and $\dim{(\mathcal{H})}=10$, respectively) and sizes of the operator pool ($\Delta=145$, $\Delta=129$ and $\Delta=49$, respectively) representative of all $p$-shell isotopes. 

All panels show $N_\text{op}$, the number of total operations that each method requires to converge the ground state, for each reference state, $\ket{ v_\alpha }$. We stress that these reference states, displayed in the horizontal axes, are all symmetry-conserving Slater determinants of the system. The blue circles correspond to the central values for our UCC simulations, with the corresponding bars indicating standard deviations, computed over different operator orderings and parameter initializations.

Figure \ref{fig:v_perf_UCC_vs_ADAPT} highlights that, while the performance of the UCC method may depend on the nucleus, the average number of total operations that it needs to converge (blue circles) is roughly independent of the reference state. For the simulation of $^{8}$Be, the average number of total operations ranges between $\langle N_\text{op}^\text{UCC} \rangle\approx(5-6)\times10^{5}$, while for $^{8}$B we find $\langle N_\text{op}^\text{UCC} \rangle\approx(4-5)\times10^{5}$ and for $^{10}$Li,  $\langle N_\text{op}^\text{UCC} \rangle\approx(4-5)\times10^{4}$. Results not shown here in the interest of brevity indicate that using a random reference state, the results are similar. 

For all these nuclei, the differences in UCC central values are much smaller than the corresponding standard deviations, indicating more sensitivity to the ordering of the operators of the pool as well as to the initial parameters of the UCC \emph{ansatz} than to the reference state. This conclusion is consistent with the analysis of Ref.~\cite{Oriel_kiss}, and also applies to the other $p$-shell nuclei we have studied with the UCC method.

\begin{figure*}[t]
    \centering
    \includegraphics[width=\textwidth]{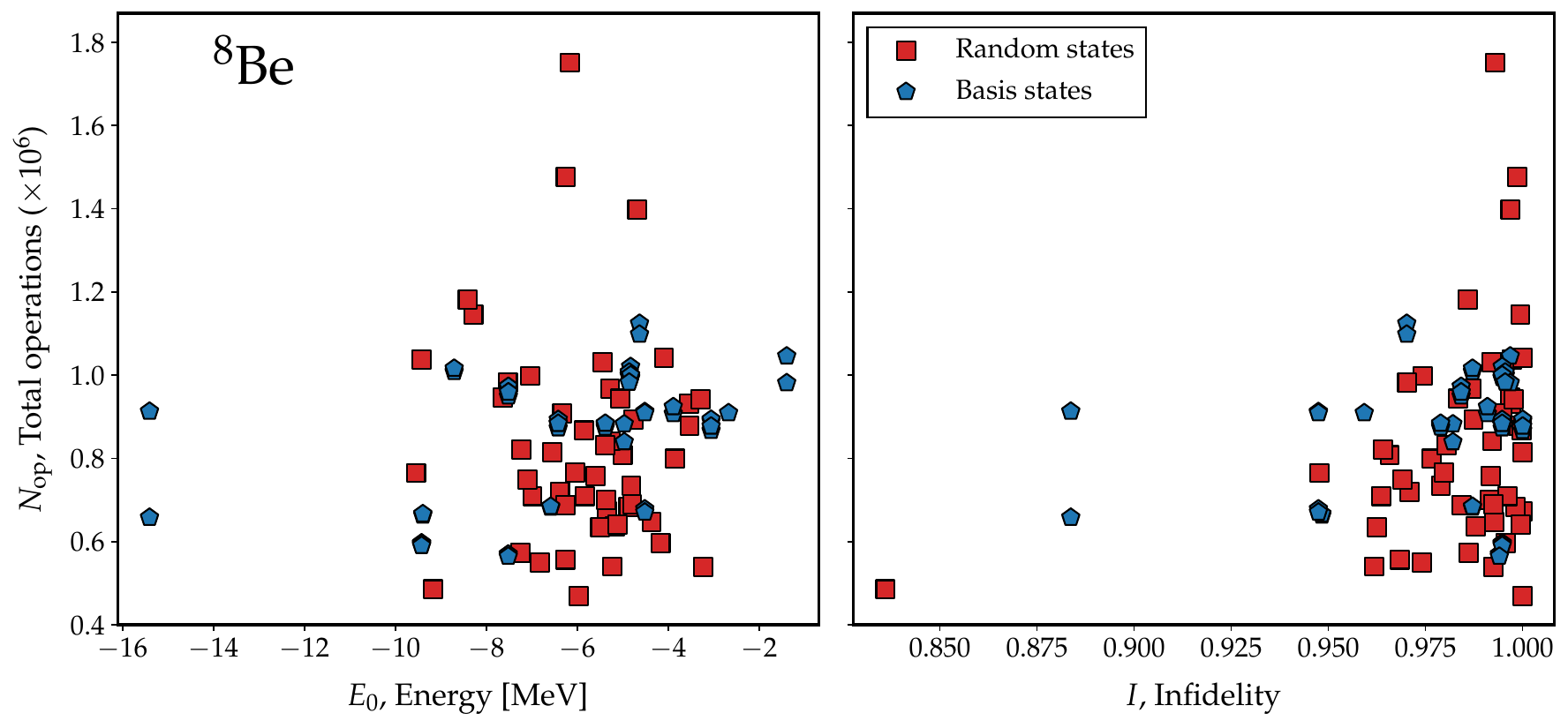}
    \caption{Total operations, $N_{\text{op}}^{\text{ADAPT}}$,needed to converge the ground state of $^{8}$Be using random states (red squares) and Slater determinants (blue pentagons) as reference states, plotted as a function of their energy, $E_0$ (left panel); and their infidelity, $I$ (right panel), with respect to the NSM ground state with energy $E=-30.29$ MeV.}
    \label{fig:Adapt_energy_overlap}
\end{figure*}

In contrast, the ADAPT simulations  in Fig.~\ref{fig:v_perf_UCC_vs_ADAPT} (orange pentagons)  indicate that, for this VQE, different reference states can lead to a different performance as measured by total operations. For instance, for $^{8}$Be, in panel (a), $N_\text{op}^\text{ADAPT}\approx(6-12)\times10^{5}$, depending on the state of the many-body basis that is used as reference state. Therefore, for some reference states (e.g. $v_{11}$ and $v_{45}$), the ADAPT minimization may reach the ground state  after twice as many total operations than the most efficient ones (e.g. $v_{28}$ and $v_{36}$). Likewise, for $^{8}$B, panel (b) indicates that the range of operations needed for ADAPT to converge presents a similarly wide variation, $N_\text{op}^\text{ADAPT}\approx(1-2)\times10^{5}$. Only for the simpler calculation of $^{10}$Li, ADAPT converges uniformly after $N_\text{op}^\text{ADAPT}\approx (2400-2700)$ total operations, regardless of the reference state.  

In conclusion, the comparison of $N_\text{op}$ between the two VQEs yields mixed results for the nuclei shown in Fig.~\ref{fig:v_perf_UCC_vs_ADAPT}. While for $^{8}$Be, the ADAPT performance only approaches the average one of UCC for the more favorable reference states, in both $^{8}$B and $^{10}$Li ADAPT reaches the ground state after fewer total operations. We defer a comprehensive discussion of these aspects to Sec.~\ref{sec:final_comparison}.  

We may exploit the dependence on reference-state choice in ADAPT simulations to try to minimize the number of total operations by choosing an adequate reference state. Establishing a clear criterion would also facilitate comparison with UCC.
Naively, one can expect that reference states that resemble the ground state would converge after fewer operations, since the minimization starts {\it closer} to the final state. We quantify this by studying the energy of the reference state, $E_{0}$, as well as the infidelity of the reference state with respect to the NSM ground state,
\begin{equation}
    I=1-|\braket{\psi_{\rm{NSM}}| v_\alpha}|^2,
    \label{eq:infidelity}
\end{equation}
which requires knowing the actual ground state beforehand. Here, we assume that a reference state with lower infidelity can be identified using prior knowledge about the system. 

We focus on $^{8}\text{Be}$, which features a significant dependence on the reference state, as seen in Figure \ref{fig:v_perf_UCC_vs_ADAPT} (a). As indicated in Table~\ref{tab: all_nuclei}, this system has a relatively large many-body dimension and a maximal operator pool size. Figure~\ref{fig:Adapt_energy_overlap} shows the number of total operations that ADAPT needs to converge the ground state as a function of the energy (left panel) and infidelity with respect to the NSM ground state (right panel) for 101 reference states: the $51$ Slater determinants forming the $^{8}\text{Be}$ many-body basis (blue pentagons), along with $50$ random states (red squares). Some many-body basis states appear indistinguishable in the panels due to their similar performance, which arises from symmetric single-particle configurations.

Figure \ref{fig:Adapt_energy_overlap} recovers the dependence on the reference state observed in Fig.~\ref{fig:v_perf_UCC_vs_ADAPT} (a): the spread in the number of operations of the many-body states $\ket{v_\alpha}$ ranges as $N_{ \text{op}}^\text{ADAPT} \approx (6-12) \times 10^5$, about a factor of two. This dependence is even larger for random reference states, where the range extends to $N_{ \text{op}}^\text{ADAPT} \approx (5-18) \times 10^5$. In other words, the most efficient state in terms of overall performance requires almost four times fewer total operations than the least efficient one. We stress that both the least and the most efficient states are random states, rather than basis states.

\begin{figure*}[t]
    \centering
    \includegraphics[width=\textwidth]{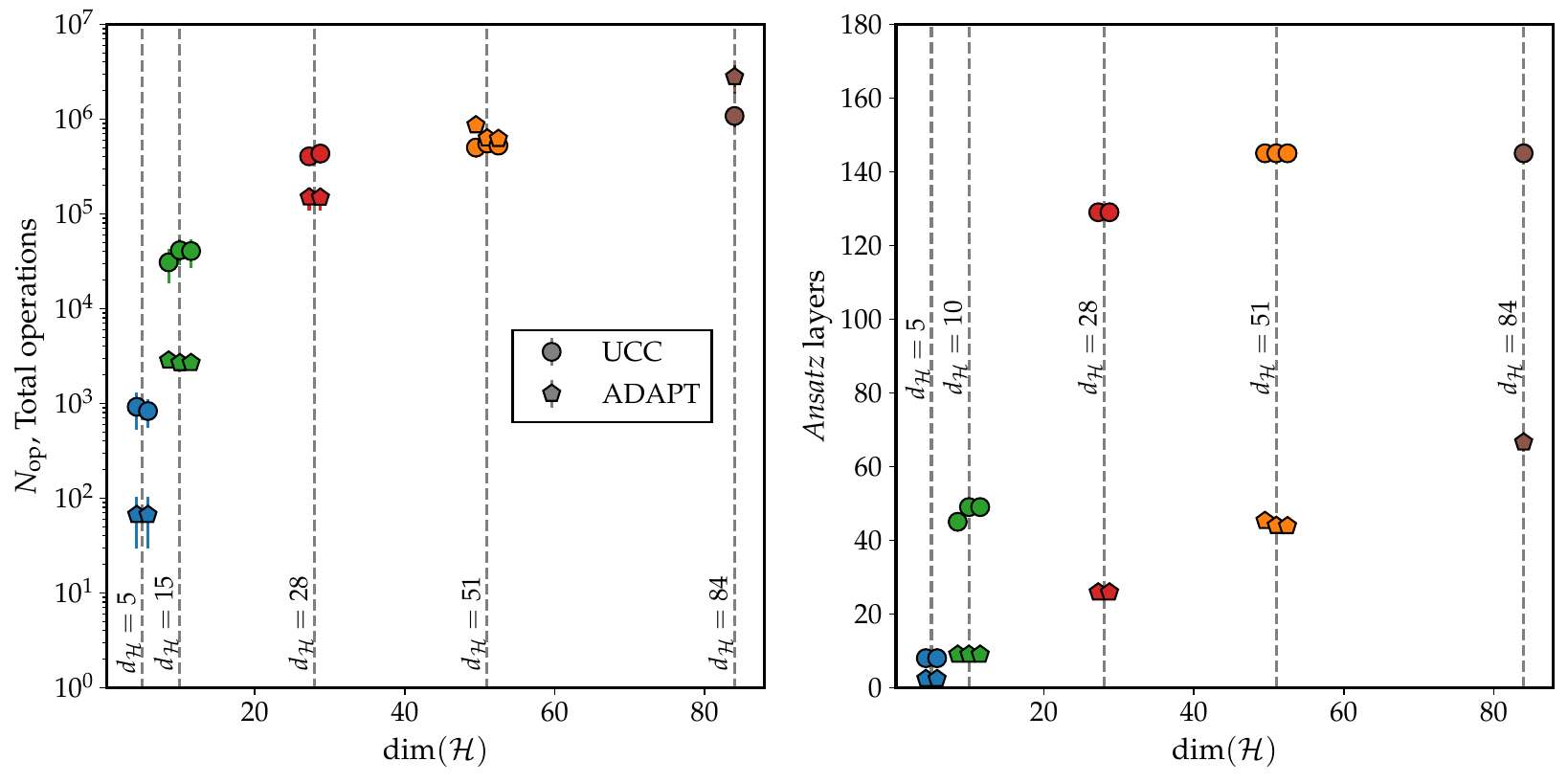}
    \caption{Average total operations (left panel) and \emph{ansatz} layers (right panel) needed to converge the ground state of all nuclei considered in this work with the ADAPT (pentagons) and UCC (circles) VQEs as a function of the dimension of the many-body basis, $\text{dim}(\mathcal{H})$. The ADAPT results average over using as reference state each state of the many-body basis, while the UCC ones average over several runs varying the pool-operator ordering and the initial parameter values, but keeping as reference state the first Slater determinant of the many-body basis.}
    \label{fig:UCC_ADAPT_all_nuclei}
\end{figure*}

Figure \ref{fig:Adapt_energy_overlap} shows that most reference states cluster around $E_0 \approx -6$~MeV and $I\approx0.95-1.0$, and need about $N_{ \text{op}}^\text{ADAPT} \approx (0.5-1.0)\times 10^6 $ total operations. While several minimizations lie outside this region and a few reference states lead to faster convergence, Fig.~\ref{fig:Adapt_energy_overlap} does not exhibit a clear pattern to identify them. Indeed, some of the most efficient reference states approach full infidelity with respect to the NSM ground state, as exemplified by the red square at the bottom right corner of the right panel of Fig.~\ref{fig:Adapt_energy_overlap}. This efficient performance is, nonetheless, similar to the one associated with the reference state with lowest infidelity, shown at the bottom left corner of the same panel. Likewise, the left panel of Fig.~\ref{fig:Adapt_energy_overlap} shows that the reference states with lowest energy---two Slater determinants---require more total operations than at least a dozen other reference states.

These findings are indicative of all other $p$-shell nuclei listed in Table~\ref{tab: all_nuclei}. They suggest that the path chosen by the ADAPT algorithm is not straightforward: reference states with similar energies or infidelities may lead to minimizations requiring quite different number of total operations. Therefore, we are not able to establish a criterion for choosing optimal ADAPT reference states, at least for states with large energy differences and high infidelities with respect to the actual ground state. In particular, given the \emph{cluster}-like shapes in Fig.~\ref{fig:Adapt_energy_overlap}, reference states with slightly lower infidelity or lower energy do not necessarily lead to an improved performance, as long as this cluster is far from the NSM exact solution with zero infidelity. Moreover, unbiased random states are not good starting points for ADAPT, since they may perform poorly, as indicated by the three red squares at the top of each panel in Fig.~\ref{fig:Adapt_energy_overlap}. 

In other words, the Slater determinants that form the many-body basis are better suited as reference states in both UCC and ADAPT. Let us now look into the generic performance of UCC and ADAPT across a whole set of representative nuclei in the $p$ shell.

\subsection{Overall performance of UCC vs ADAPT in $p$-shell nuclei}\label{sec:final_comparison}

Finally, we turn towards a more global perspective and compare the performance of the UCC and ADAPT VQEs across all $p$-shell nuclei in Table \ref{tab: all_nuclei}. The left panel of Fig.~\ref{fig:UCC_ADAPT_all_nuclei} shows the number of total operations needed to obtain the ground state as a function of the dimension of the Hilbert space for these isotopes. 
For UCC, the mean values $\langle N^\text{UCC}_\text{op} \rangle$ (solid circles) and the corresponding standard deviation, are computed over different operator orderings and initial parameter values, as specified in Sec.~\ref{sec:Results}. In these results, and given the uniformity observed in Fig.~\ref{fig:v_perf_UCC_vs_ADAPT}, we employ as a reference state the first state of the many-body basis. 

For ADAPT, in contrast, the mean values $\langle N^\text{ADAPT}_\text{op} \rangle$ (solid pentagons) are computed by averaging over all the $\text{dim}(\mathcal{H})$ possible reference states for each isotope. Figure~\ref{fig:UCC_ADAPT_all_nuclei} shows central values as well as standard deviations, although the latter appear small on logarithmic scale. 

The left panel of Fig.~\ref{fig:UCC_ADAPT_all_nuclei} highlights the most relevant results of our work. As expected, the number of total operations grows with the dimension of the Hilbert space for both VQEs. However, this scaling is milder for UCC than for ADAPT. For instance, in the simplest nuclei with $\text{dim}(\mathcal{H})<51$, ADAPT needs fewer operations to converge to the ground state than UCC. More specifically, for the lightest systems---$^6$Be and $^6$He---with $\text{dim}(\mathcal{H})=5$, ADAPT converges with $\langle N^\text{ADAPT}_\text{op} \rangle<100$, whereas $\langle N^\text{UCC}_\text{op} \rangle\approx1000$. Likewise, for $^6$Li, $^{10}$Li and $^{10}$N, with $\text{dim}(\mathcal{H})=10$, ADAPT simulations need $\langle N^\text{ADAPT}_\text{op} \rangle \approx 3000$, as opposed to $\langle N^\text{UCC}_\text{op} \rangle>12000$. For $\text{dim}(\mathcal{H})=28$, corresponding to $^8$Li and $^8$B, again ADAPT requires fewer total operations, $\langle N^\text{ADAPT}_\text{op} \rangle\approx 2 \times 10^5$ compared to $\langle N^\text{UCC}_\text{op} \rangle\approx 5 \times 10^5$.  This difference between VQEs, however, disappears as we reach $\dim(\mathcal{H})=51$. Indeed, for these nuclei---$^8$Be, $^{10}$Be and $^{10}$C---the UCC method slightly outperforms ADAPT in terms of total operations. Finally, for $^{10}\text{B}$, which has the largest Hilbert space of the $p$ shell, $\dim(\mathcal{H})=84$, UCC, with $\langle N^\text{UCC}_\text{op} \rangle\approx10^6$, presents a clear advantage compared to ADAPT, that requires $\langle N^\text{ADAPT}_\text{op} \rangle \approx 3 \times 10^6$.

In order to understand the different scaling of the ADAPT and UCC VQEs with the dimension of the Hilbert space, the right panel of Fig.~\ref{fig:UCC_ADAPT_all_nuclei} focuses on the number of layers of the \emph{ansatz} used in the minimization. For all UCC runs, this corresponds to the dimension of the energy surface, equal to the operator pool size $\Delta$, listed in Table~\ref{tab: all_nuclei}. In contrast, for ADAPT the number of layers is smaller, as not all pool operators are used in the minimization. Although this number depends on the reference state used on each ADAPT minimization, the variance is not appreciable. The right panel of Fig.~\ref{fig:UCC_ADAPT_all_nuclei} shows that the number of ADAPT layers grows linearly with the dimension of the Hilbert space, which is consistent with Ref.~\cite{ADAPT_ICCUB}. In contrast, for UCC the number of layers grows faster for simple nuclei but later it stabilizes at $\Delta=145$. This limit is set by the number of two-body excitation operators in the $p$ shell that preserve the total magnetic number $M$. This is, once this limit is reached, which in the $p$ shell occurs for a relatively small Hilbert-space dimension, the depth of the UCC \emph{ansatz} does not increase any further, in contrast to the linear scaling observed for ADAPT. As a result, the relative number of UCC \emph{ansatz} layers decreases, from more than five times the maximum number of ADAPT ones when $\dim(\mathcal{H})= 28$, to just about twice as many for $\dim(\mathcal{H})= 84$.

This behavior leads to a better overall performance of UCC for the most complex $p$-shell nuclei, as illustrated in the left panel of Fig.~\ref{fig:UCC_ADAPT_all_nuclei}. The reason is that, in contrast to UCC, ADAPT requires multiple iterations, each with an additional computational cost. As the \emph{ansatz} becomes more complex, each minimization increases the  number of total operations needed for each evaluation. Ultimately, this causes ADAPT to require considerably more quantum operations compared to UCC for mid-shell nuclei. 

\section{Conclusions and outlook}
\label{sec:Conclusions}
In conclusion, we simulate the ground state energy of several nuclei in the $p$ shell with relative energy errors $\varepsilon\leq 10^{-4}$ with respect to classical NSM benchmarks. We examine both the UCC and ADAPT algorithms for the same systems under identical conditions. In order to compare the two VQEs, we propose a metric to measure the number of computations carried out by the quantum device, the total operations, $N_\text{op}$. In our analysis, we avoid state-dependent conclusions by averaging the results obtained using several reference states, including randomized ones. For UCC, we also average over simulations using different initial parameters. 

The performance of the UCC method does not depend on the choice of reference state. In contrast, when using ADAPT, the total operations needed to reach the ground state can vary by over a factor of two. However, we do not find any pattern to identify the most suitable reference states, at least as long as their energy and infidelity with respect to the actual ground state are considerably high. Specifically, random states $   \ket{  v_\text{rand} }= \sum_\alpha c_\alpha \ket{ v_\alpha }\, ,$ may perform particularly poorly. Therefore, the Slater determinants that form the many-body basis, emerge as the more suitable choice for reference states for both VQEs.

We compare the average number of total operations of both UCC and ADAPT, simulating eleven nuclei across the $p$ shell. As expected, the number of total operations required to converge the ground state increases with the Hilbert space dimension, but the scaling for UCC is milder than for ADAPT. For $^6$He, $^6$Be, $^{6,8,10}$Li, $^{10}$N and $^8$B, which are relatively close to a shell closure (thus with a small many-body space dimension, $\dim(\mathcal{H})<51$), ADAPT is more efficient than UCC, because it requires much fewer \emph{ansatz} layers to simulate the ground state. On the other hand, for nuclei closer to the mid-shell such as $^{8,10}$Be, $^{10}$C and $^{10}$B (with $\dim (\mathcal{H}) \geq 51$), UCC outperforms ADAPT in terms of total operations. This is because for these systems the relative number of layers is more comparable between UCC and ADAPT. Furthermore, ADAPT performs additional quantum operations in each iteration that end up exceeding the cost of the UCC in a single, but more complex, minimization. 

While a detailed future study is warranted, we expect this general behavior to hold in NSM simulations beyond the $p$ shell. Typically, ADAPT simulations may require fewer quantum resources for systems near closed shells, where the ground state may converge without a full $\Delta$-dimensional energy minimization, which makes UCC simulations expensive. However, toward the mid-shell, the additional operations required by several ADAPT iterations may render this method more costly than UCC implemented with a deeper circuit. Assuming this behavior, it would be very interesting to explore the transition of the VQE of choice from ADAPT to UCC---which occurs at $\dim(\mathcal{H})= 51$ in the $p$ shell---in other configuration spaces.  

Indeed, we plan to extend the current analysis to nuclei in the $sd$ and $pf$ shells. This would increase the number of many-body basis states and the operator pool size, thereby enlarging the dimensions of the energy landscapes. Such an extension would allow us to explore the scaling of the UCC and ADAPT methods as a function of the dimension of the system. Besides, we also aim to simulate the error profile of the associated quantum hardware. This aspect may impact both VQEs with a different loss in precision, as they demand varying numbers of total operations and circuit depth to simulate the same nuclear ground states. Such study would help us to identify the most suitable VQE for investigating nuclear structure with state-of-the-art quantum devices.

\acknowledgments

We thank Xavier Roca Maza for useful discussions.
This work is financially supported by 
MCIN/AEI/10.13039/501100011033 from the following grants: PID2020-118758GB-I00 and PID2023-147112NB-C22; 
RYC-2017-22781 and RYC2018-026072 through the “Ram\'on
y Cajal” program funded by FSE “El FSE invierte en tu futuro”; 
CNS2022-135529 and CNS2022-135716 funded by the
“European Union NextGenerationEU/PRTR”, and 
and CEX2024-001451-M  to the “Unit of Excellence Mar\'ia de Maeztu 2025-2031” award to the Institute of Cosmos Sciences; and by the Generalitat de Catalunya, through grant 2021SGR01095. 
This work is also financially supported by the Ministry of Economic Affairs and Digital Transformation of the Spanish Government through the QUANTUM ENIA project call – Quantum Spain project, and by the European Union through the Recovery, Transformation and
Resilience Plan – NextGenerationEU within the framework of the Digital Spain 2026 Agenda.

\bibliography{biblio}

\end{document}